\documentclass[aps,pra,onecolumn,notitlepage,superscriptaddress]{revtex4-1}

\usepackage{graphicx}
\usepackage{natbib}
\usepackage[breaklinks]{hyperref}
\usepackage{amsmath}
\usepackage{mathtools}

\newcommand{\hflev}[4]{\textit{#1}$_{#2/#3}$, \textit{F}=#4}

\begin{document}

\title{Nonlinear photon-atom coupling with 4Pi microscopy} 
\author{Yue-Sum Chin}
\affiliation{Centre for Quantum Technologies, 3 Science Drive 2, Singapore 117543}
\author{Matthias Steiner}
\affiliation{Centre for Quantum Technologies, 3 Science Drive 2, Singapore 117543}
\affiliation{Department of Physics, National University of Singapore, 2 Science Drive 3, Singapore 117542}
\author{Christian Kurtsiefer}
\affiliation{Centre for Quantum Technologies, 3 Science Drive 2, Singapore 117543}
\affiliation{Department of Physics, National University of Singapore, 2 Science Drive 3, Singapore 117542}
\email[]{}
\date{\today}

%\begin{abstract}

%\end{abstract}

\maketitle
\textbf{ 
Implementing nonlinear interactions between single photons and single atoms is at the forefront of optical physics.
Motivated by the prospects of deterministic all-optical quantum logic, many efforts are currently underway to find suitable experimental techniques~\cite{Tiecke2014,Shomroni2014,Hacker2016}. 
Focusing the incident photons onto the atom with a lens yielded promising results~\cite{Wineland1987,Vamivakas2007,Tey2008,Wrigge2008,Piro2011,Maser2016}, but is limited by diffraction to moderate interaction strengths~\cite{Tey2009}.  
However, techniques to exceed the diffraction limit are known from high-resolution imaging.
Here, we adapt a super-resolution imaging technique, 4Pi microscopy~\cite{Hell1992}, to efficiently couple light to a single atom. 
We observe 36.6(3)\% extinction of the incident field, and a modified photon statistics of the transmitted field -- indicating nonlinear interaction at the single-photon level. 
Our results pave the way to few-photon nonlinear optics with individual atoms in free space. 
  }

\begin{figure}[b]
  \centering
\includegraphics[width=.7\columnwidth]{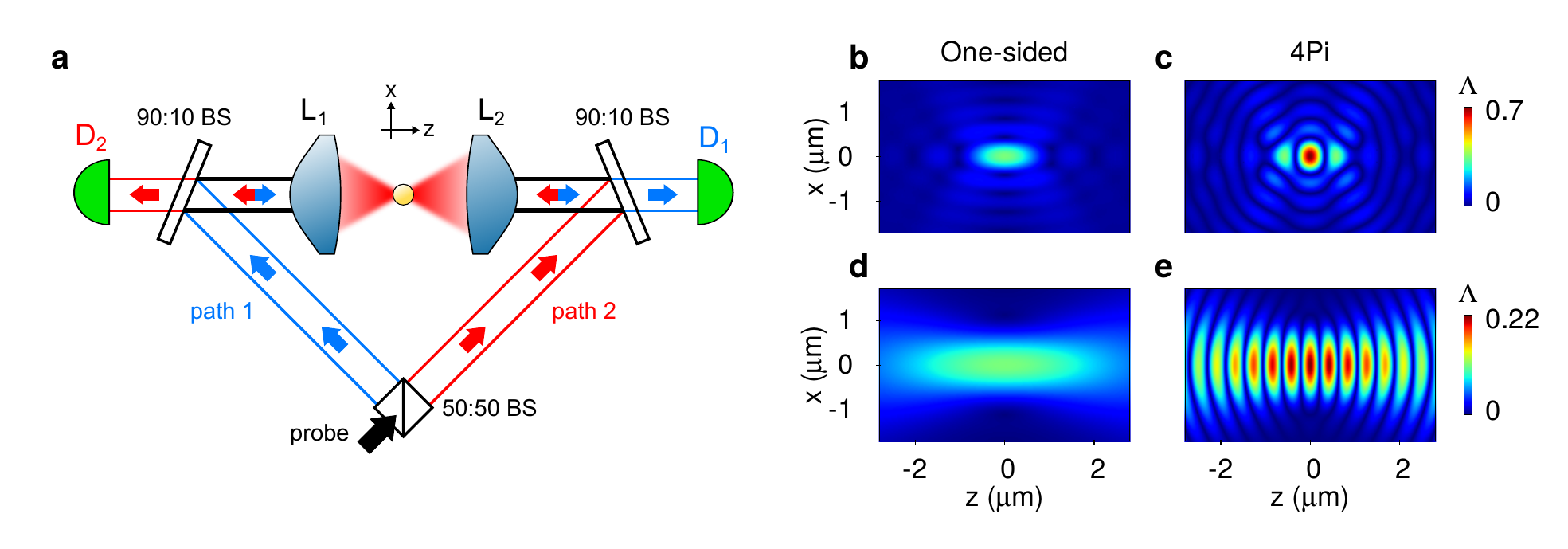}
    \caption{\label{fig:figure1}
    \textbf{Concept of 4Pi illumination.}
    \textbf{a}, Schematics of the optical setup. 
    The probe beam~(black arrow) is split into path 1 (blue arrows) and path 2 (red arrows). 
    The two beams then illuminate the atom from counter-propagating directions. 
    Asymmetric beamsplitters are used to sample the probe light after passing the atom. 
    The probe light in path~1(2) is coupled into a single mode fiber connected to detector~$D_{1(2)}$.  
    By blocking one path, we recover the commonly employed one-sided illumination.
    BS:\,beam splitter, $L_{1(2)}$:\,high numerical aperture lens, $D_{1(2)}$:\,avalanche photodetector. 
    \textbf{b-e}, Numerical results of the coupling efficiency~$\Lambda$ near the focal point considering a Gaussian field resonantly driving a circularly polarized dipole transition near 780\,nm~\cite{Tey2009}. 
    The field is assumed to constructively interfere at the focal point for the 4Pi configuration. 
    \textbf{b/c}, Focusing parameters corresponding to an objective with numerical aperture NA$=0.95$. 
    \textbf{d/e}, Focusing parameters used in this experiment~(input beam waist~$w_0=2.7\,$mm at lens, focal length~$f=5.95$\,mm). 
 }
\end{figure}

\noindent
To realize nonlinear interactions between a few propagating photons and a single atom in free space, the photons need to be tightly focused to a small volume~\cite{Sondermann2007,Leuchs2012}. 
From high-resolution imaging it is well-known that a small focal volume requires optical elements which cover a large fraction of the solid angle~\cite{Abbe1873}. 
While standard confocal optical microscopy accomplished already very small probe volumes, the excitation light is focused through a lens that can cover only up to half of the solid angle, limiting the axial resolution due to a focal volume elongated along the optical axis.
This limitation has been overcome by using two opposing lenses with coinciding focal points, known as 4Pi arrangement~\cite{Hell1992}:   
The path of the incident beam is split, and the object is coherently illuminated by two counter-propagating parts of the field simultaneously~(Fig.~\ref{fig:figure1}a). 
In this way the input mode covers almost the entire solid angle, limited only by the numerical aperture of the focusing lenses. 
The symmetry between imaging and excitation of quantum emitter suggests that a 4Pi arrangement can also be used to efficiently couple light to an atom.  
This intuitive argument is confirmed by numerical simulations of the electric field distribution near the focal point, from which we obtain the spatial mode overlap of the atomic dipole mode with the input mode, referred to as the light-atom coupling efficiency~$\Lambda= |E_\textrm{focus}|^2/|E_\textrm{max}|^2$, where $E_\textrm{focus(max)}$ is the (maximally possible) amplitude of the incident electric field component parallel to the atomic dipole~(Fig.~\ref{fig:figure1}b-f)~\cite{Tey2009,Golla2012}. 

In our experiment, we hold a single $^{87}$Rb atom between two lenses with a far off-resonant optical dipole trap (FORT) operating at a wavelength 851\,nm~\cite{Schlosser2001}. 
We compare 4Pi and one-sided illumination by performing a transmission experiment with a weak coherent field driving the closed transition~\mbox{5\hflev{S}{1}{2}{2}, $m_F$=-2} to  \mbox{5\hflev{P}{3}{2}{3}, $m_F$=-3} near 780\,nm~\cite{Chin2017}. 
The probe beam originates from a collimated output of a single mode fiber. 
After splitting into path~1 and path~2, the beam is focused onto the atom through lenses~$L_{1}$ and $L_{2}$~(see Fig.~\ref{fig:figure1}a). 
The opposing lens re-collimates the probe beam, which is then via an asymmetric beam splitter coupled into a single mode fiber connected to avalanche photodetector~$D_1$ or $D_2$, respectively~(see Supplementary Information for details).  
The electric fields at the detectors are superpositions of the probe field and the field scattered by the atom. 
To derive the total electric field, we adapt the theoretical description of Ref.~\cite{Tey2009,Slodifmmodeheckclsecika2010} to account for the contributions of the two counter-propagating probe fields. 
The optical power~$P_1$ at detector~$D_1$ depends then on the power in the individual beam paths~$P_\textrm{1(2),in}$ and the light-atom coupling efficiency~$\Lambda_{1(2)}$ of path~1(2), 
\begin{equation}\label{eq:P_i}
P_\textrm{1} =  \left( \sqrt{P_\textrm{1,in}} - 2\Lambda_1 \sqrt{P_\textrm{1,in}} - 2 \sqrt{\Lambda_1 \Lambda_2}  \sqrt{P_\textrm{2,in}}\right)^2,
\end{equation}
where we assume that the two fields interfere constructively at the focal point. 
Similarly, the power at detector~$D_2$ is obtained by exchanging subscripts $1\leftrightarrow2$.  
From equation~\ref{eq:P_i} we obtain the expected values for
 the individual transmission $T_\textrm{1(2)} = P_{1(2)}/P_\textrm{1(2),in}$, and the total transmission $T_\textrm{total} = (P_\textrm{1}+ P_\textrm{2})/ (P_\textrm{1,in}+ P_\textrm{2,in})$. 
For example, for a one-sided illumination through lens~$L_1$, i.e. $P_\textrm{2,in}=0$, the transmission measured at detector~$D_1$ takes the well known expression $T_1  =   \left(1- 2\Lambda_1\right)^2$~\cite{Tey2009,Slodifmmodeheckclsecika2010}. 
In the 4Pi configuration, we determine the total coupling~$\Lambda_\textrm{total}$ from the total transmission $T_\textrm{total}= \left(1- 2\Lambda_\textrm{total}\right)^2$. 
From equation~\ref{eq:P_i} we find that the power splitting~$P_\textrm{2,in}=P_\textrm{1,in} \Lambda_1/\Lambda_2$ optimizes the total coupling to~$\Lambda_\textrm{total}=\Lambda_1+\Lambda_2$.

\begin{figure}
  \centering
  \includegraphics[width=0.7\columnwidth]{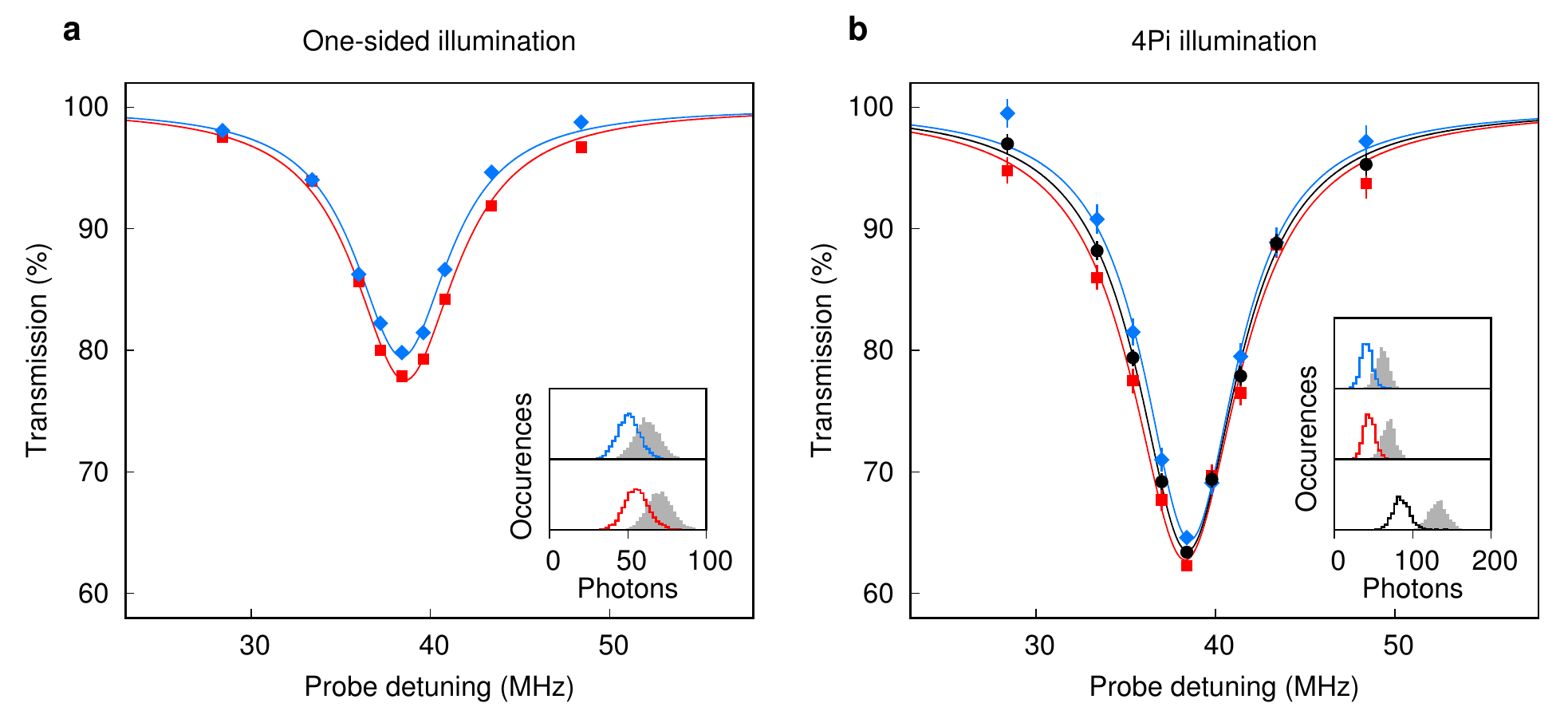}
    \caption{\label{fig:histo_spec} 
    \textbf{ Extinction of a weak coherent probe beam.} 
    \textbf{a}, One-sided illumination via path~1~(blue diamonds) or path~2~(red squares). 
    Solid lines are Lorentzian fits. 
    The inset shows the normalized histogram of detected photons during the probe cycle (solid line) and reference cycle (gray) for the resonant data point. 
    \textbf{b}, Same as \textbf{a} but with 4Pi illumination. 
    The total transmission (black circles) is obtained from the sum of detectors~$D_1$ and $D_2$. 
    Error bars represent one standard deviation of propagated Poissonian counting uncertainties. 
    The FORT shifts the resonance frequency by approximately $38.5\,$MHz compared to the natural transition frequency. 
}
\end{figure}

Figure~\ref{fig:histo_spec}a shows the transmission spectrum of a weak coherent field for one-sided illumination, either via path 1 (blue) or path 2 (red). 
Comparing the resonant transmission $T_1= 77.9(2)\%$ and $T_2= 79.8(3)\%$ to equation~\ref{eq:P_i}, we find similar coupling efficiencies, $\Lambda_1 = 0.059(1)$ and~$\Lambda_2 = 0.053(1)$, as expected for our symmetric arrangement with two nominally identical lenses. 
Therefore, the maximum coupling expected in the 4Pi configuration is~$\Lambda_\textrm{total}=\Lambda_1+\Lambda_2= 0.112(4)$, assuming perfect phase matching of the fields and ideal positioning of the atom.

In the 4Pi configuration the atom needs to be precisely placed at an anti-node of the incident field~(Fig.~\ref{fig:figure1}e). 
To this end, we tightly confine the atom along the optical axis with an additional blue-detuned standing wave dipole trap~(761\,nm). 
As the atom is loaded probabilistically into the optical lattice, we use a simple postselection technique to check whether the atom is trapped close to an anti-node of the incident field~(see Methods).
Figure~\ref{fig:histo_spec}b shows the observed transmission when the atom is illuminated in the 4Pi arrangement. 
The increased light-atom coupling is evident from the strong reduction of transmission. 
The individual transmissions $T_\textrm{1}=62.3(5)\%$, $T_\textrm{2}=64.6(5)\%$, and the total transmission $T_\textrm{total} = 63.4(3)\%$ are significantly lower compared to the one-sided illumination.
The corresponding total coupling of $\Lambda_\textrm{total}=0.102(1)$ is close to the theoretical prediction of~$0.112(4)$. 

\begin{figure}
  \centering
  \includegraphics[width=0.35\columnwidth]{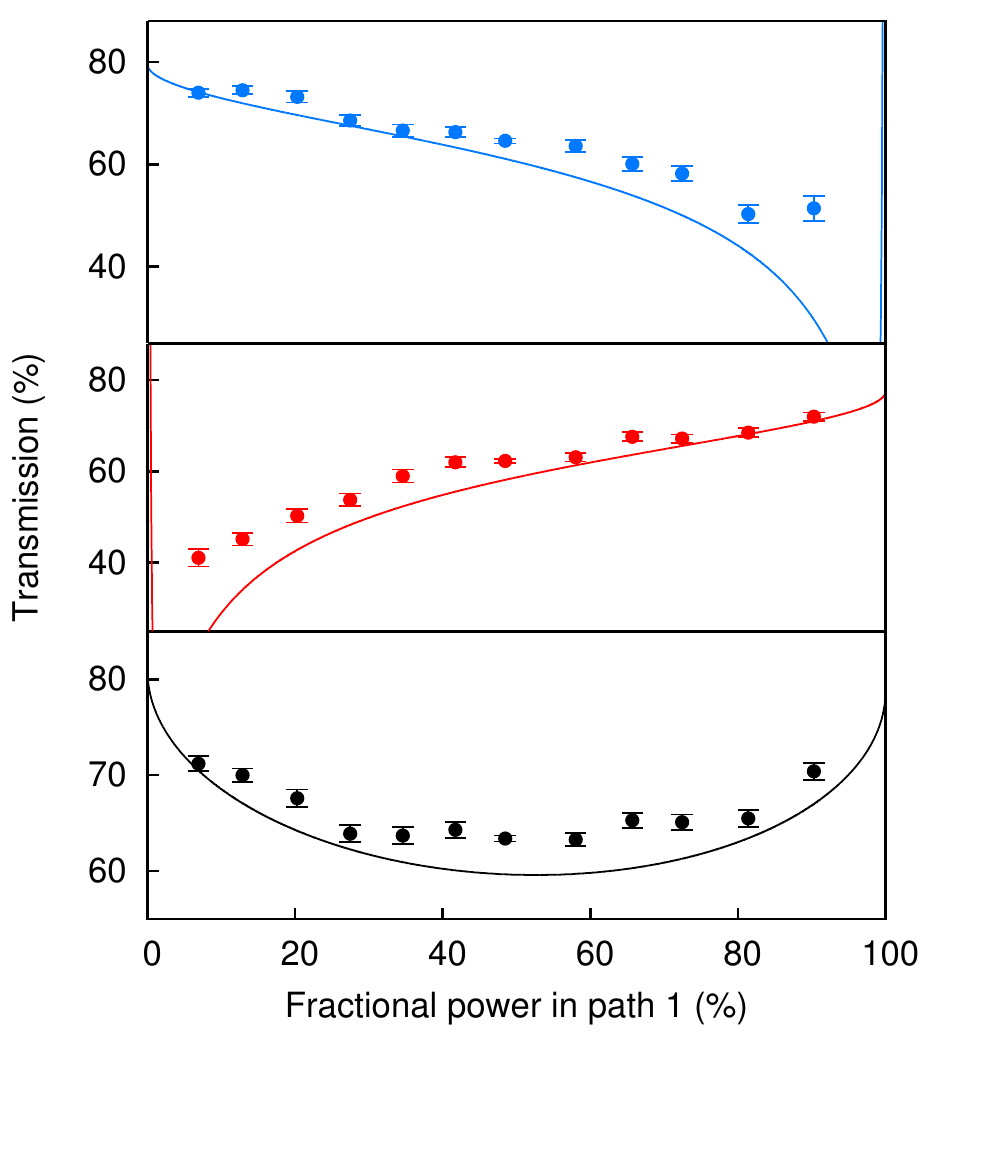}
    \caption{\label{fig:ratio}
    \textbf{Resonant transmission for different power splittings between path~1 and path~2.}
    Transmission at detector $D_1$ (top), $D_2$ (center) and the total transmission $D_1+D_2$ (bottom). 
    The total number of incident photons is kept constant. 
    Solid lines are  $T_\textrm{1(2)}$ and $T_\textrm{total}$ derived from equation~\ref{eq:P_i}. 
    Error bars represent one standard deviation of propagated Poissonian counting uncertainties. 
 }
\end{figure}

We next show that for a symmetric arrangement~$\Lambda_1\approx\Lambda_2$, the highest interaction is achieved with an equal power splitting $P_\textrm{2,in}\approx P_\textrm{1,in}$. 
Figure~\ref{fig:ratio} displays the resonant transmissions for different relative beam power in the two paths.  
For imbalanced beam power, the total transmission is increased, albeit with a fairly weak dependence. 
In contrast, we find a strong dependence of the individual transmissions on the relative beam power: 
For $P_\textrm{1,in}\approx 12 P_\textrm{2,in}$, the total transmission is
still low, $T_\textrm{total} =71.2(8)\%$, but the two values for the
individual transmissions are no longer equal: $T_\textrm{1,4Pi}=74.0(8)\%$,
$T_\textrm{2,4Pi}=41(2)\%$, 
in qualitative agreement with equation~\ref{eq:P_i}~(solid lines in Fig.~\ref{fig:ratio}).

\begin{figure}
  \centering
  \includegraphics[width=0.7\columnwidth]{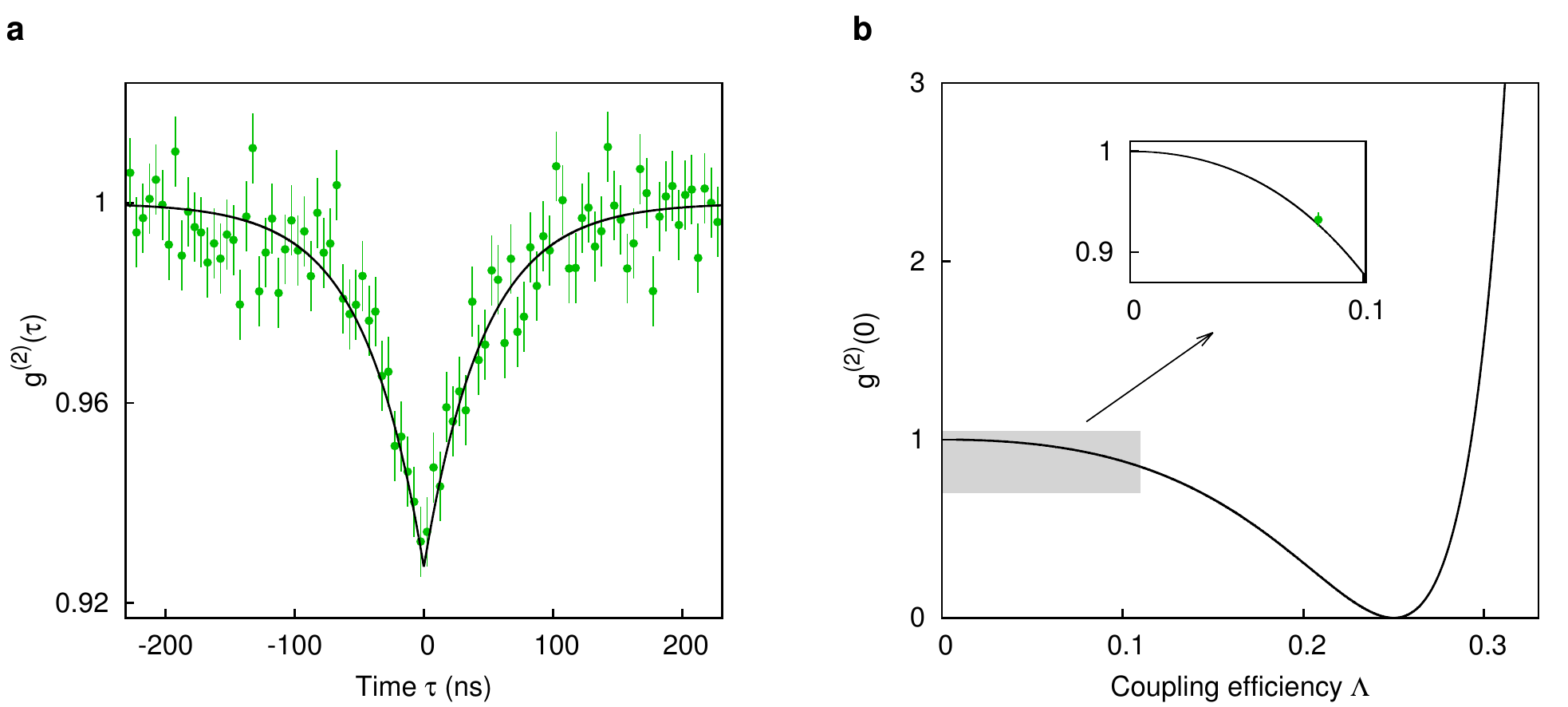}
    \caption{\label{fig:g2}
    \textbf{Modified photon statistics due to nonlinear interaction. }
    \textbf{a}, Intensity correlation of transmitted light with a time bin width of 5\,ns.  
    Solid line is the theoretical prediction without free parameter~(see equation~\ref{eq:g2}).
    \textbf{b}, Dependence on the coupling efficiency~$\Lambda$. 
    The inset is a zoom into the region of our data point for clarity, and the 
    solid line is $g^{(2)}(0)$ from equation~\ref{eq:g2}. 
 }
\end{figure}

The nonlinear character of the photon-atom interaction can induce effective attractive or repulsive interactions between two photons~\cite{Shen2007}.
These interactions can be observed as modification of the photon statistics of the transmitted field if the initial field contains multi-photon contributions~\cite{Birnbaum2005,Dayan2008,Reinhard2012,Hoi2012,Peyronel2012}. 
For a weak coherent driving field,
the second-order correlation function~$g^{(2)}(\tau)$ takes the specific form~\cite{Chang2007,Zheng2010} 
\begin{equation}\label{eq:g2}
 g^{(2)}(\tau) = e^{-\Gamma_0 \tau} \left( \left(\frac{2\Lambda}{1-2\Lambda}\right)^2 - e^{\frac{\Gamma_0 \tau}{2}} \right)^2,
\end{equation}
where $\Gamma_0=2\pi \times 6.07\,$MHz is the excited state linewidth. 
By time-tagging the detection events at detector~$D_1$ and $D_2$ during the probe phase, we obtain $g^{(2)}(\tau)= \langle p_1(t) p_2(t+\tau)\rangle / ( \langle p_1(t)\rangle \langle p_2(t+\tau)\rangle)$, where $p_{1(2)}(t)$ is the detection probability at detector~$D_{1(2)}$ at time~$t$, and $\langle \rangle$ denotes the long time average. 
To acquire sufficient statistics, we use $50\%$ more photons in the probe pulse as compared to Fig.~\ref{fig:histo_spec}, and also atoms which are not optimally coupled to the probe field~(see Methods). 
From the resulting average transmission~$T_\textrm{total} =70.3(3)\%$, we deduce an average coupling~$\Lambda_\textrm{total}=0.0808(5)$ for this experiment.
As shown in Fig.~\ref{fig:g2}, we find a clear signature of nonlinear photon-atom interaction in the intensity correlations of the transmitted light. 
The observed photon anti-bunching~$g^{(2)}(0) = 0.934(7)\%$ is in good agreement with equation~\ref{eq:g2}. 
Here, for fair comparison with equation~\ref{eq:g2}, we account for a small photon bunching effect~($\approx1.7\%$, see Methods) due to the diffusive atomic motion~\cite{Gomer1998,Weber2006}. 
For stronger light-atom coupling the changes of the photon statistics are expected to be more significant~(Fig.~\ref{fig:g2}b). 
Notably, for $\Lambda =0.25$ the transmitted and the reflected light show
anti-bunching ($g^{(2)}(0) =0$), that means the atom acts as a photon turnstile and converts a coherent field completely into a single photon field. 
The transmission for this light-atom coupling is~$T_\textrm{total}=25\%$ (see equation~\ref{eq:P_i}). 
Photon bunching ($g^{(2)}(0) > 1$) for large values of $\Lambda$ signals an enhanced probability for multiple photons to be transmitted, essentially because the atom cannot scatter multiple photons simultaneously.

Our work establishes the 4Pi arrangement as an effective technique to couple a propagating field to an atom. 
This opens exciting prospects to implement effective interactions between photons with tightly focused free space modes and single atoms. 
Strongly interacting photons could find application in imaging, metrology, quantum computing and cryptography, and constitute a novel platform to study many-body physics~\cite{Chang2008,Chang2014}. 
The presented approach forms an experimental alternative to waveguide/cavity quantum electrodynamics~\cite{Birnbaum2005,Reiserer2015} and Rydberg quantum optics~\cite{Pritchard2010,Peyronel2012,Firstenberg2016}.  
While the achieved nonlinearity of the photon-atom interaction, observed as modification of the photon statistics, does not create strongly correlated photons yet, the 4Pi arrangement eases the technical requirements to the focusing lens considerably, making the implementation of strong photon-photon interaction feasible. 
In the near future, we expect that by using higher numerical aperture lenses, the 4Pi arrangement will allow the efficient conversion of a coherent beam into single photons.  

\section*{Methods}
\subsection*{Experimental sequence and postselection of the atom position}
The experimental sequence starts with loading a single atom from a cold ensemble in a magneto-optical trap  into a far-off resonant dipole trap. 
Once trapped, the atom undergoes molasses cooling for 5\,ms. 
We then apply a bias magnetic field of 0.74\,mT along the optical axis and optically pump the atom into the 5\hflev{S}{1}{2}{2}, $m_F$=-2 state. 
Subsequently, we perform two transmission experiments during which we switch on the probe field for 1\,ms each. 
The first transmission measurement is used to determine the light-atom coupling~$\Lambda$, the second one to check whether the atom is trapped at an anti-node of probe field. 
To obtain the relative transmission, we also detect the instantaneous probe power for each transmission experiment by optically pumping the atom into the \mbox{5\hflev{S}{1}{2}{1}} hyperfine state, which shifts the atom out of resonance with the probe field by 6.8\,GHz, and reapply the probe field. 

The postselection of the atom position is performed as follows:
We select the detection events in the first transmission experiment conditioned on the number of photons detected in the second one. 
The frequency of the probe field during the second transmission experiment is set to be resonant with the atomic transition. 
For the data shown in Fig.~\ref{fig:histo_spec}b and Fig.~\ref{fig:ratio} we use a threshold which selects approximately 0.5\% of the total events as a trade-off between data acquisition rate and selectiveness of the atomic position.
To measure the second-order correlation function of the transmitted light (Fig.~\ref{fig:g2}a),  we choose a higher threshold which selects $\sim$10\% of the experimental cycles. 

\subsection*{Normalization of second-order correlation function}
We measure the second order correlation function  of the transmitted light using detector~$D_1$ and $D_2$ as the two detectors of a Hanbury-Brown and Twiss setup.
The photodetection events are time tagged during the probe phase, and sorted into a time delay histogram. 
We obtain the normalized correlation function~$g^{(2)}(\tau)$ by dividing the number of occurrences by $r_1 r_2 \Delta t T$, where $r_{1(2)}$ is the mean count rate at detector $D_{1(2)}$, $\Delta t$ is the time bin width, and $T$ is the total measurement time. 
For times $100\,$ns $< \tau < 1\,\mu$s, we find super-Poissonian intensity
correlations $g^{(2)}(\tau)>1$, which are induced by the atomic motion through the trap. 
Although the amplitude of the correlations is small, we nevertheless perform a deconvolution for a better comparison to Eq.~\ref{eq:g2}. 
The correlations are expected to decay exponentially for diffusive motion,
thus we fit  $f(\tau)= 1 + a_0 \exp{(-\tau/\tau_d)}$ to $g^{(2)}(\tau)$,
resulting in $\tau_d=0.71(8)\,\mu$s and $a_0=0.019(2)$.
Figure~\ref{fig:g2} shows the second order correlation function after deconvolution of the diffusive motion, i.e., after division by $f(\tau)$~(see Supplementary Information).

%\clearpage

\section*{SUPPLEMENTARY INFORMATION}

\subsection*{Optical setup}

\begin{figure}[h]
\centering
  \includegraphics[width=0.9\columnwidth]{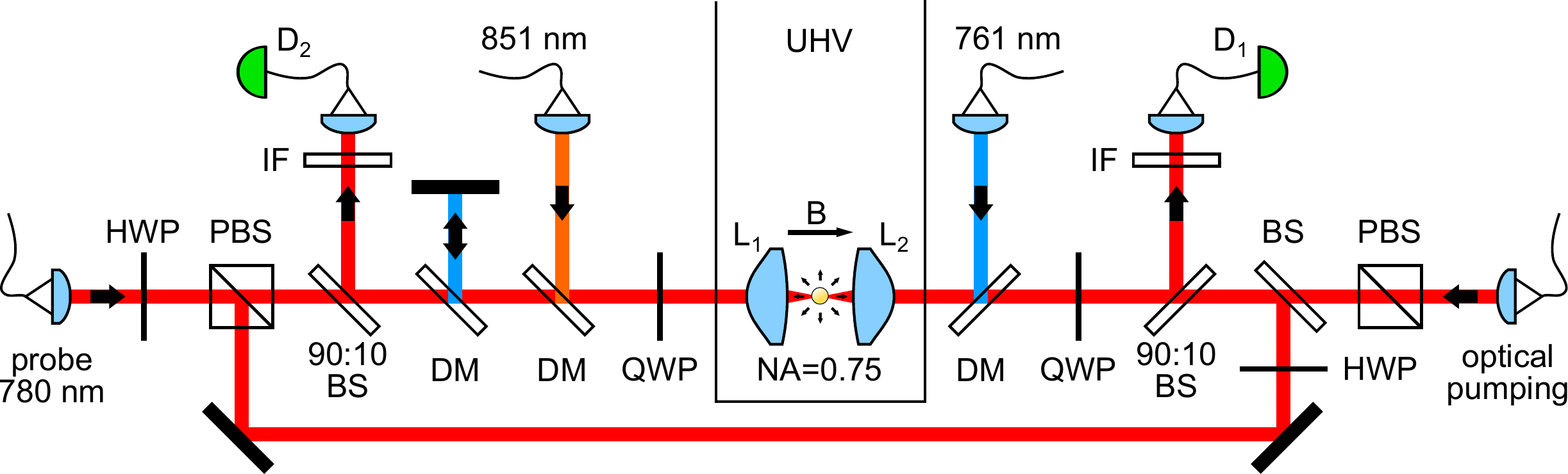}
  \caption{\label{fig:setup}\textbf{Optical setup.} D$_\textrm{1}$, D$_\textrm{2}$:~avalanche photodetectors (APDs), IF:~interference filter, HWP, QWP:~half- and quarter-wave plates, (P)BS:~(polarizing) beam splitter, DM:~dichroic mirror, L$_\textrm{1}$, L$_\textrm{2}$:~high numerical aperture lenses, B:~magnetic field, UHV:~ultra-high vacuum chamber.
 }
\end{figure}

\textit{Probe path.}
The Gaussian probe beam is delivered from a single-mode fiber, collimated and split into two paths~(Fig.~\ref{fig:setup}). 
The power ratio in the two paths is controlled by a half-wave plate and a polarizing beam splitter. 
Half- and quarter-wave plates ensure the same polarization~($\sigma^-$) in both paths at the position of the atom. 
After passing through the lens pair, the probe light is coupled into single mode fibers connected to avalanche photodetectors. 
We optimize the fiber couplings to collect the probe light and measure 40\% coupling loss that is due to imperfect mode matching. 

\textit{Dipole traps.} 
We trap single $^{87}$Rb atoms with a red-detuned far-off-resonant dipole trap (FORT) at 851\,nm. 
The circularly polarized ($\sigma^+$) beam is focused to a waist~$w_0\approx1.4\,\mu$m, which results in a trap depth of~$U_0 = k_B \times 1.88$\,mK.
The position of the trap is adjusted to maximize the collected atomic fluorescence at the detectors $D_1$ and $D_2$. 
In addition, we use a blue-detuned FORT at 761\,nm in standing wave configuration overlapping with the red-detuned FORT to increase the axial confinement. 
The blue-detuned FORT is linearly polarized and has a trap depth of approximately $0.1$\,mK along the optical axis. 

\subsection*{Experimental sequence and postselection of atom position}
\textit{Measurement strategy.}
To fully utilize the 4Pi arrangement the atom needs to placed at an anti-node of the probe field. 
Unfortunately, the interference pattern of the probe field changes over time owing to slow drifts in the optical path lengths. 
The probe-atom coupling is further affected by similar drifts of the optical lattice, and the probabilistic loading into particular lattice sites. 
Here we exploit that once an atom is loaded, the timescale for a transmission experiment is much shorter (milliseconds) than the timescale of the drifts~(minutes). 
Therefore, each experimental cycle consists of two independent transmission experiments: one to check whether the atom is trapped at the right position and one to determine the light-atom interaction. 
In the actual sequence we first perform the light-atom interaction experiment before checking the atom position. 
In this way we minimize the effect of recoil heating from the probe field. 

\begin{figure}[h]
\centering
  \includegraphics[width=0.55\columnwidth]{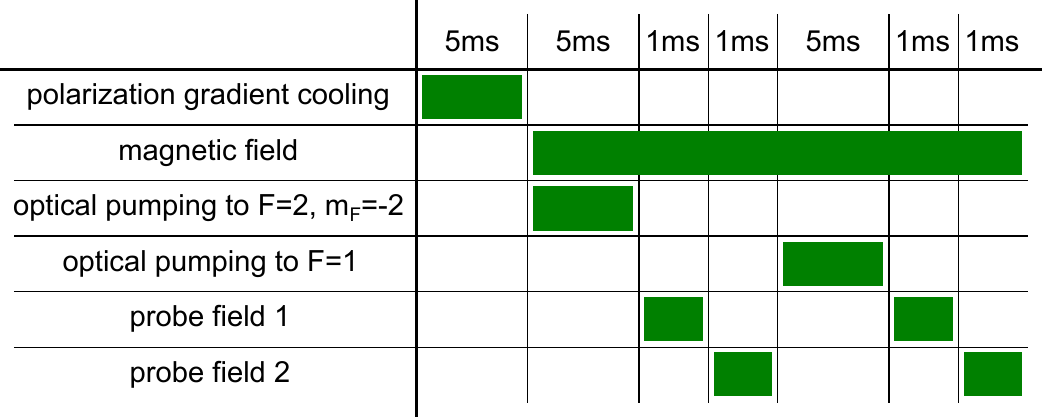}
  \caption{\label{fig:sequence}\textbf{Experimental sequence.} }
\end{figure}

\textit{Experimental sequence.}
The experiment begins upon the loading of a single atom. 
We then perform polarization gradient cooling for 5\,ms~(Fig.~\ref{fig:sequence}), which
cools the atom to a temperature of about 16\,$\mu$K. 
A bias magnetic field of~$0.74$\,mT is applied along the optical axis, and the atom is prepared in the 5\hflev{S}{1}{2}{2}, $m_F$=-2 state by optical pumping. 
Next, two probe fields are applied each for 1\,ms, separated by a $4\,\mu$s pause.
We tune the frequency of the first probe, for example, to obtain the transmission spectra shown in Fig.~\ref{fig:histo_spec}. 
The second probe cycle is used to check whether the atom has been trapped at an anti-node of the probe field.
For this, the frequency of the probe field is set to be resonant with the atomic transition. 
Subsequently, we perform a reference measurement to obtain the instantaneous probe power. 
We first optically pump the atom to the \mbox{5\hflev{S}{1}{2}{1}} hyperfine
state, shifting the atom out of resonance with the probe field by 6.8\,GHz,
after which we reapply the two probe fields. 
The detection events at avalanche photodetectors~$D_\textrm{1}$ and $D_\textrm{2}$ are recorded during all probe cycles. 

\begin{figure}[ht]
\centering
  \includegraphics[width=0.7\columnwidth]{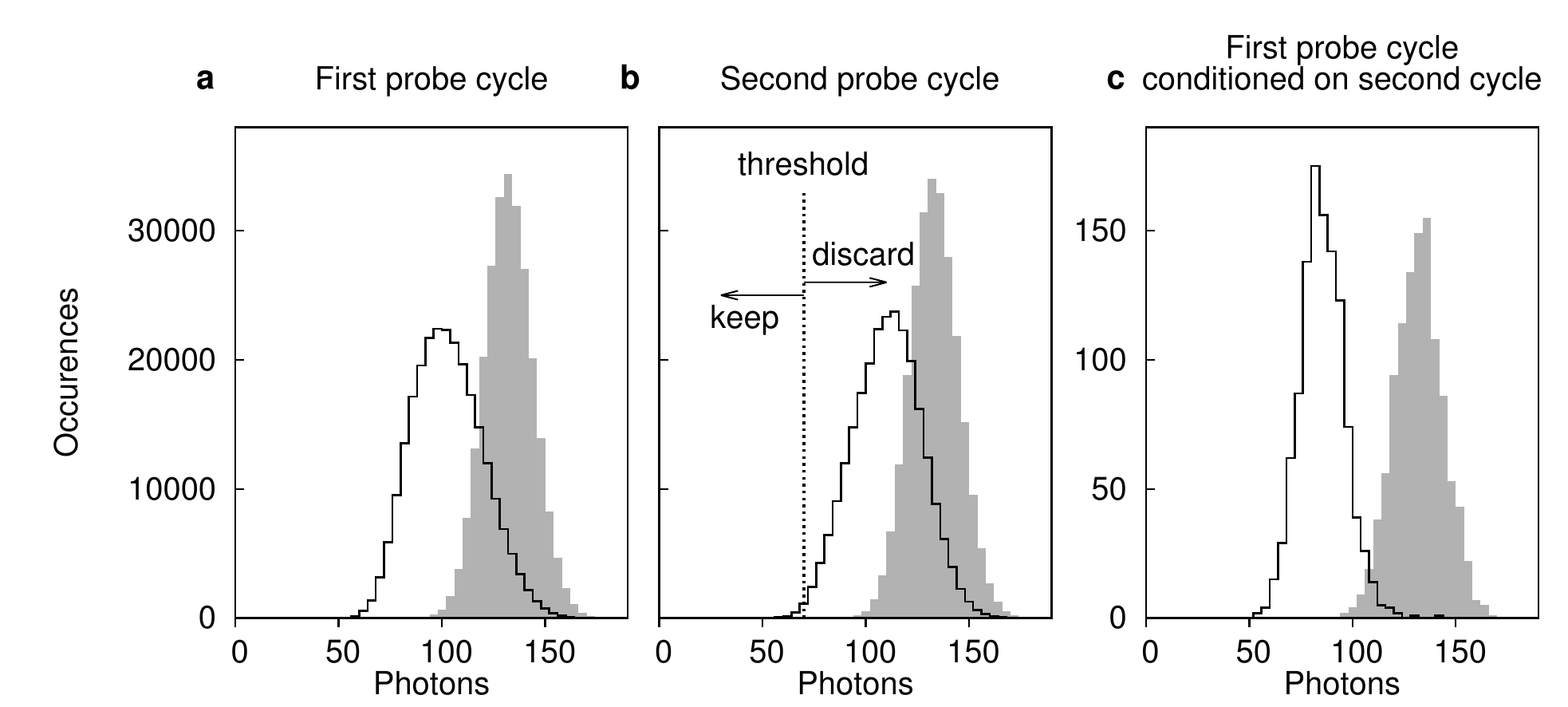}
  \caption{\label{fig:hist}\textbf{Postselection of atom position.} Photon counting histogram recorded during probe (solid line) and reference (gray) cycle.  
		    The total number of detected photons is computed as the sum of detectors $D_1$ and $D_2$. 
		    \textbf{a}, First probe cycle for the case when the probe field is resonant to the atomic transition. 
		    \textbf{b}, Second probe cycle. 
		    The dotted line marks the set threshold for a postselection of approximately 0.5\% of the total events. 
		    \textbf{c}, Resultant events of the first probe cycle conditioned on the second cycle using the marked threshold in \textbf{b}. 
		    }
\end{figure}

\textit{Postselection of atom position.}
We illustrate the postselection procedure for the case in which the probe field during the first probe cycle is resonant with the atomic transition. 
Figure~\ref{fig:hist}a/b shows the histogram of detected photons in the first/second probe cycle.  
The position of the atom is postselected based on the detected transmission during the second probe cycle.  
For an atom loaded into a desired site of the potential well, the transmission is low. 
Hence, we discard detection events in the first probe cycle if the number of photons detected in the second cycle is above a threshold value.   
Figure~\ref{fig:hist}c shows the histogram of detected photons in the first probe cycle after postselection. 
For the transmission measurements shown in Fig.~\ref{fig:histo_spec} and Fig.~\ref{fig:ratio}, we use a photocount threshold that selects
approximately $0.5\%$ of the total events, trading off between data acquisition rate and selectiveness of the atomic position. 
For the case of one-sided illumination, this postselection procedure does not change the observed transmission. 
In the second order correlation measurement, we use a higher threshold value to speed up the data acquisition, selecting $10\%$ of the total events. 
The correlations shown in Fig.~\ref{fig:g2} are the result of approximately 200 hours of measurement time.

\subsection*{Photon statistics of transmitted light}
\begin{figure}[h]
\centering
  \includegraphics[width=0.7\columnwidth]{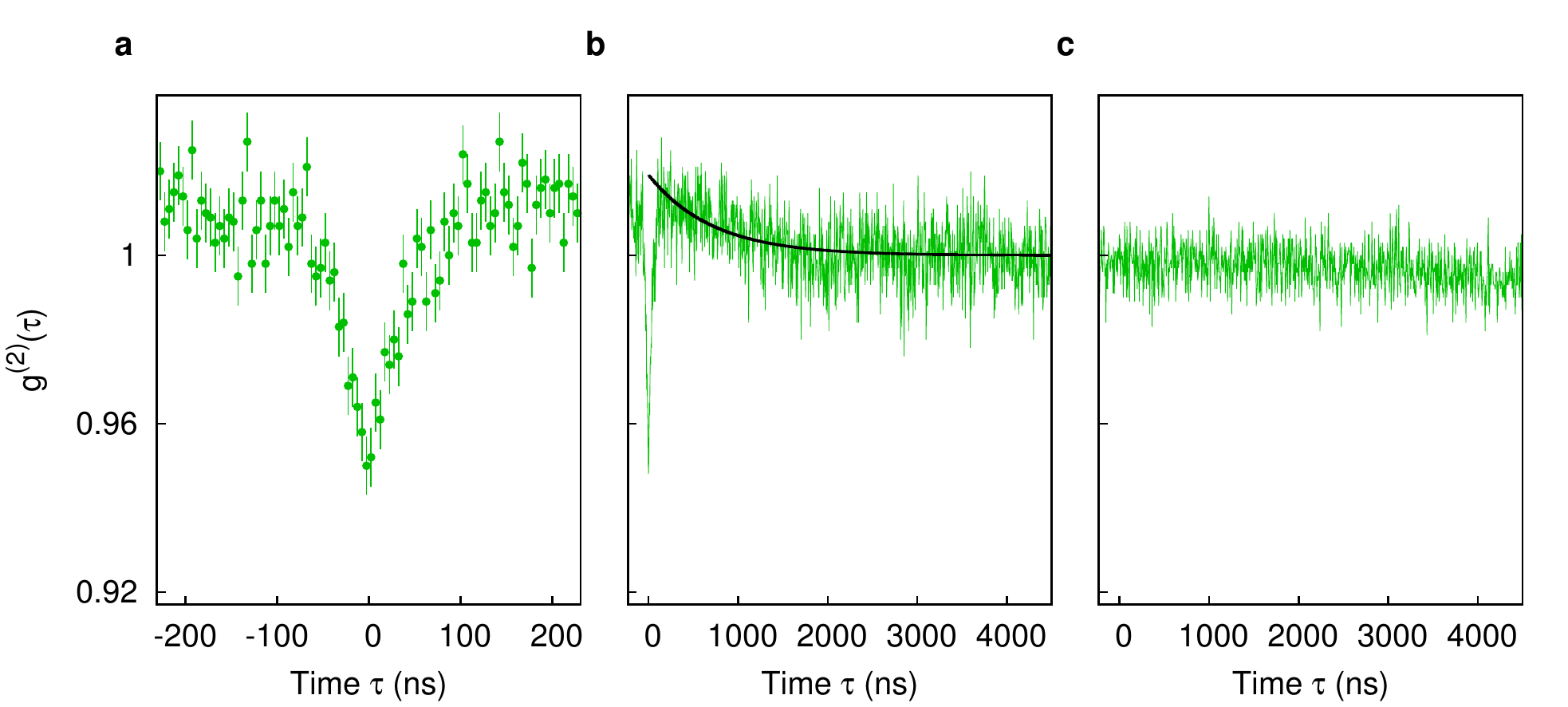}
  \caption{ \label{fig:g2raw}\textbf{Photon bunching due to atomic motion. }
		    \textbf{a,} Normalized second order correlation function without deconvolution of the diffusive atomic motion. 
		    \textbf{b,} Same as \textbf{a} but with extended range. 
		    Solid line is a fit to  $f(\tau)= 1 + a_0 \exp{(-\tau/\tau_d)}$ with $a_0=0.019(2)$, $\tau_d=0.71(8)\,\mu$s and a reduced $\chi^2=1.07$. 
		    \textbf{c,} Same as \textbf{b} but computed from events during the reference cycle, i.e., when the atom is not resonant with probe field.
		    }
\end{figure}

\textit{Normalized second order correlation function.}
We compute the second order correlation function from the time-tagged photodetection events at detector~$D_1$ and $D_2$. 
We sort the photodetection events into a time delay histogram and obtain the normalized correlation function by dividing the number of occurrences by $r_1 r_2 \Delta t T$, where $r_{1(2)}$ is the mean count rate at detector $D_{1(2)}$, $\Delta t$ is the time bin width and $T$ is the total measurement time. 
To make the normalization robust against intensity drifts of the probe power, we perform the normalization for every  1\,ms-long measurement cycle, obtaining the normalized correlation function~$g^{(2)}_i(\tau)$~(index $i$ describes the measurement cycle) and then $g^{(2)}(\tau)$ from the  weighted mean
\begin{equation}\label{eq:g2norm}
 g^{(2)}(\tau) =  \frac{\sum^{N}_{i=1} g^{(2)}_i(\tau)  (r_{1,i}+r_{2,i})}{\sum^{N}_{i=1} (r_{1,i}+r_{2,i})}.
\end{equation}
Figure~\ref{fig:g2raw}a-b shows $g^{(2)}(\tau)$ around $\tau=0$ and for longer time delays. 
For large $\tau$, the correlation disappears, and $g^{(2)}(\tau)$ approaches unity. 
However, for $100\,$ns $< \tau < 1\,\mu$s, $g^{(2)}(\tau)$ shows super-Poissonian intensity correlations $g^{(2)}(\tau)>1$. 
Similar correlations have been observed in the fluorescence of single atoms in dipole traps induced by the atomic motion through the trap~(Ref.[27,28]). 

\textit{Deconvolution of the diffusive atomic motion.}
Although the amplitude of the correlations is small, we nevertheless perform a deconvolution for a better comparison to Eq.~2.  
For diffusive motion the correlations are expected to decay exponentially,
thus we fit  $f(\tau)= 1 + a_0 \exp{(-\tau/\tau_d)}$ to $g^{(2)}(\tau)$,
resulting in $a_0=0.019(2)$, $\tau_d=0.71(8)\,\mu$s, with a reduced $\chi^2=1.07$~(Figure~\ref{fig:g2raw}b, black solid line). 
We note that the timescale $\tau_d$ of these correlations is much larger than the excited state lifetime~$1/\Gamma_0=26.2\,$ns. 
Figure~\ref{fig:g2} shows the second order correlation function corrected for the diffusive motion, i.e. after division by $f(\tau)$. 
No additional correlations are present in the transmitted light during the
reference cycle, i.e., when the atom is not resonant with probe field (Fig.~\ref{fig:g2raw}c). 

\begin{acknowledgments}
We acknowledge the support of this work by the Ministry of Education in
Singapore (AcRF Tier 1) and the National Research Foundation, Prime Minister's
office.
M.\,Steiner acknowledges support by the Lee Kuan Yew Postdoctoral Fellowship.
\end{acknowledgments}

\bibliographystyle{apsrev4-1}
%\bibliography{fourpi.bib}{}

\begin{thebibliography}{33}%
\makeatletter
\providecommand \@ifxundefined [1]{%
 \@ifx{#1\undefined}
}%
\providecommand \@ifnum [1]{%
 \ifnum #1\expandafter \@firstoftwo
 \else \expandafter \@secondoftwo
 \fi
}%
\providecommand \@ifx [1]{%
 \ifx #1\expandafter \@firstoftwo
 \else \expandafter \@secondoftwo
 \fi
}%
\providecommand \natexlab [1]{#1}%
\providecommand \enquote  [1]{``#1''}%
\providecommand \bibnamefont  [1]{#1}%
\providecommand \bibfnamefont [1]{#1}%
\providecommand \citenamefont [1]{#1}%
\providecommand \href@noop [0]{\@secondoftwo}%
\providecommand \href [0]{\begingroup \@sanitize@url \@href}%
\providecommand \@href[1]{\@@startlink{#1}\@@href}%
\providecommand \@@href[1]{\endgroup#1\@@endlink}%
\providecommand \@sanitize@url [0]{\catcode `\\12\catcode `\$12\catcode
  `\&12\catcode `\#12\catcode `\^12\catcode `\_12\catcode `\%12\relax}%
\providecommand \@@startlink[1]{}%
\providecommand \@@endlink[0]{}%
\providecommand \url  [0]{\begingroup\@sanitize@url \@url }%
\providecommand \@url [1]{\endgroup\@href {#1}{\urlprefix }}%
\providecommand \urlprefix  [0]{URL }%
\providecommand \Eprint [0]{\href }%
\providecommand \doibase [0]{http://dx.doi.org/}%
\providecommand \selectlanguage [0]{\@gobble}%
\providecommand \bibinfo  [0]{\@secondoftwo}%
\providecommand \bibfield  [0]{\@secondoftwo}%
\providecommand \translation [1]{[#1]}%
\providecommand \BibitemOpen [0]{}%
\providecommand \bibitemStop [0]{}%
\providecommand \bibitemNoStop [0]{.\EOS\space}%
\providecommand \EOS [0]{\spacefactor3000\relax}%
\providecommand \BibitemShut  [1]{\csname bibitem#1\endcsname}%
\let\auto@bib@innerbib\@empty
%</preamble>
\bibitem [{\citenamefont {Tiecke}\ \emph {et~al.}(2014)\citenamefont {Tiecke},
  \citenamefont {Thompson}, \citenamefont {de~Leon}, \citenamefont {Liu},
  \citenamefont {Vuletic},\ and\ \citenamefont {Lukin}}]{Tiecke2014}%
  \BibitemOpen
  \bibfield  {author} {\bibinfo {author} {\bibfnamefont {T.~G.}\ \bibnamefont
  {Tiecke}}, \bibinfo {author} {\bibfnamefont {J.~D.}\ \bibnamefont
  {Thompson}}, \bibinfo {author} {\bibfnamefont {N.~P.}\ \bibnamefont
  {de~Leon}}, \bibinfo {author} {\bibfnamefont {L.~R.}\ \bibnamefont {Liu}},
  \bibinfo {author} {\bibfnamefont {V.}~\bibnamefont {Vuletic}}, \ and\
  \bibinfo {author} {\bibfnamefont {M.~D.}\ \bibnamefont {Lukin}},\ }\href
  {http://dx.doi.org/10.1038/nature13188} {\bibfield  {journal} {\bibinfo
  {journal} {Nature}\ }\textbf {\bibinfo {volume} {508}},\ \bibinfo {pages}
  {241} (\bibinfo {year} {2014})}\BibitemShut {NoStop}%
\bibitem [{\citenamefont {Shomroni}\ \emph {et~al.}(2014)\citenamefont
  {Shomroni}, \citenamefont {Rosenblum}, \citenamefont {Lovsky}, \citenamefont
  {Bechler}, \citenamefont {Guendelman},\ and\ \citenamefont
  {Dayan}}]{Shomroni2014}%
  \BibitemOpen
  \bibfield  {author} {\bibinfo {author} {\bibfnamefont {I.}~\bibnamefont
  {Shomroni}}, \bibinfo {author} {\bibfnamefont {S.}~\bibnamefont {Rosenblum}},
  \bibinfo {author} {\bibfnamefont {Y.}~\bibnamefont {Lovsky}}, \bibinfo
  {author} {\bibfnamefont {O.}~\bibnamefont {Bechler}}, \bibinfo {author}
  {\bibfnamefont {G.}~\bibnamefont {Guendelman}}, \ and\ \bibinfo {author}
  {\bibfnamefont {B.}~\bibnamefont {Dayan}},\ }\href {\doibase
  10.1126/science.1254699} {\bibfield  {journal} {\bibinfo  {journal}
  {Science}\ }\textbf {\bibinfo {volume} {345}},\ \bibinfo {pages} {903}
  (\bibinfo {year} {2014})}\BibitemShut {NoStop}%
\bibitem [{\citenamefont {Hacker}\ \emph {et~al.}(2016)\citenamefont {Hacker},
  \citenamefont {Welte}, \citenamefont {Rempe},\ and\ \citenamefont
  {Ritter}}]{Hacker2016}%
  \BibitemOpen
  \bibfield  {author} {\bibinfo {author} {\bibfnamefont {B.}~\bibnamefont
  {Hacker}}, \bibinfo {author} {\bibfnamefont {S.}~\bibnamefont {Welte}},
  \bibinfo {author} {\bibfnamefont {G.}~\bibnamefont {Rempe}}, \ and\ \bibinfo
  {author} {\bibfnamefont {S.}~\bibnamefont {Ritter}},\ }\href
  {http://dx.doi.org/10.1038/nature18592} {\bibfield  {journal} {\bibinfo
  {journal} {Nature}\ }\textbf {\bibinfo {volume} {536}},\ \bibinfo {pages}
  {193} (\bibinfo {year} {2016})}\BibitemShut {NoStop}%
\bibitem [{\citenamefont {Wineland}\ \emph {et~al.}(1987)\citenamefont
  {Wineland}, \citenamefont {Itano},\ and\ \citenamefont
  {Bergquist}}]{Wineland1987}%
  \BibitemOpen
  \bibfield  {author} {\bibinfo {author} {\bibfnamefont {D.~J.}\ \bibnamefont
  {Wineland}}, \bibinfo {author} {\bibfnamefont {W.~M.}\ \bibnamefont {Itano}},
  \ and\ \bibinfo {author} {\bibfnamefont {J.~C.}\ \bibnamefont {Bergquist}},\
  }\href {\doibase 10.1364/OL.12.000389} {\bibfield  {journal} {\bibinfo
  {journal} {Opt. Lett.}\ }\textbf {\bibinfo {volume} {12}},\ \bibinfo {pages}
  {389} (\bibinfo {year} {1987})}\BibitemShut {NoStop}%
\bibitem [{\citenamefont {Vamivakas}\ \emph {et~al.}(2007)\citenamefont
  {Vamivakas}, \citenamefont {Atat\"ure}, \citenamefont {Dreiser},
  \citenamefont {Yilmaz}, \citenamefont {Badolato}, \citenamefont {Swan},
  \citenamefont {Goldberg}, \citenamefont {Imamo{\u{g}}lu},\ and\ \citenamefont
  {\"Unl\"u}}]{Vamivakas2007}%
  \BibitemOpen
  \bibfield  {author} {\bibinfo {author} {\bibfnamefont {A.~N.}\ \bibnamefont
  {Vamivakas}}, \bibinfo {author} {\bibfnamefont {M.}~\bibnamefont
  {Atat\"ure}}, \bibinfo {author} {\bibfnamefont {J.}~\bibnamefont {Dreiser}},
  \bibinfo {author} {\bibfnamefont {S.~T.}\ \bibnamefont {Yilmaz}}, \bibinfo
  {author} {\bibfnamefont {A.}~\bibnamefont {Badolato}}, \bibinfo {author}
  {\bibfnamefont {A.~K.}\ \bibnamefont {Swan}}, \bibinfo {author}
  {\bibfnamefont {B.~B.}\ \bibnamefont {Goldberg}}, \bibinfo {author}
  {\bibfnamefont {A.}~\bibnamefont {Imamo{\u{g}}lu}}, \ and\ \bibinfo {author}
  {\bibfnamefont {M.~S.}\ \bibnamefont {\"Unl\"u}},\ }\href {\doibase
  10.1021/nl0717255} {\bibfield  {journal} {\bibinfo  {journal} {Nano Letters}\
  }\textbf {\bibinfo {volume} {7}},\ \bibinfo {pages} {2892} (\bibinfo {year}
  {2007})}\BibitemShut {NoStop}%
\bibitem [{\citenamefont {Tey}\ \emph {et~al.}(2008)\citenamefont {Tey},
  \citenamefont {Chen}, \citenamefont {Aljunid}, \citenamefont {Chng},
  \citenamefont {Huber}, \citenamefont {Maslennikov},\ and\ \citenamefont
  {Kurtsiefer}}]{Tey2008}%
  \BibitemOpen
  \bibfield  {author} {\bibinfo {author} {\bibfnamefont {M.~K.}\ \bibnamefont
  {Tey}}, \bibinfo {author} {\bibfnamefont {Z.}~\bibnamefont {Chen}}, \bibinfo
  {author} {\bibfnamefont {S.~A.}\ \bibnamefont {Aljunid}}, \bibinfo {author}
  {\bibfnamefont {B.}~\bibnamefont {Chng}}, \bibinfo {author} {\bibfnamefont
  {F.}~\bibnamefont {Huber}}, \bibinfo {author} {\bibfnamefont
  {G.}~\bibnamefont {Maslennikov}}, \ and\ \bibinfo {author} {\bibfnamefont
  {C.}~\bibnamefont {Kurtsiefer}},\ }\href
  {http://dx.doi.org/10.1038/nphys1096} {\bibfield  {journal} {\bibinfo
  {journal} {Nat Phys}\ }\textbf {\bibinfo {volume} {4}},\ \bibinfo {pages}
  {924} (\bibinfo {year} {2008})}\BibitemShut {NoStop}%
\bibitem [{\citenamefont {Wrigge}\ \emph {et~al.}(2008)\citenamefont {Wrigge},
  \citenamefont {Gerhardt}, \citenamefont {Hwang}, \citenamefont {Zumofen},\
  and\ \citenamefont {Sandoghdar}}]{Wrigge2008}%
  \BibitemOpen
  \bibfield  {author} {\bibinfo {author} {\bibfnamefont {G.}~\bibnamefont
  {Wrigge}}, \bibinfo {author} {\bibfnamefont {I.}~\bibnamefont {Gerhardt}},
  \bibinfo {author} {\bibfnamefont {J.}~\bibnamefont {Hwang}}, \bibinfo
  {author} {\bibfnamefont {G.}~\bibnamefont {Zumofen}}, \ and\ \bibinfo
  {author} {\bibfnamefont {V.}~\bibnamefont {Sandoghdar}},\ }\href
  {http://dx.doi.org/10.1038/nphys812} {\bibfield  {journal} {\bibinfo
  {journal} {Nat Phys}\ }\textbf {\bibinfo {volume} {4}},\ \bibinfo {pages}
  {60} (\bibinfo {year} {2008})}\BibitemShut {NoStop}%
\bibitem [{\citenamefont {Piro}\ \emph {et~al.}(2011)\citenamefont {Piro},
  \citenamefont {Rohde}, \citenamefont {Schuck}, \citenamefont {Almendros},
  \citenamefont {Huwer}, \citenamefont {Ghosh}, \citenamefont {Haase},
  \citenamefont {Hennrich}, \citenamefont {Dubin},\ and\ \citenamefont
  {Eschner}}]{Piro2011}%
  \BibitemOpen
  \bibfield  {author} {\bibinfo {author} {\bibfnamefont {N.}~\bibnamefont
  {Piro}}, \bibinfo {author} {\bibfnamefont {F.}~\bibnamefont {Rohde}},
  \bibinfo {author} {\bibfnamefont {C.}~\bibnamefont {Schuck}}, \bibinfo
  {author} {\bibfnamefont {M.}~\bibnamefont {Almendros}}, \bibinfo {author}
  {\bibfnamefont {J.}~\bibnamefont {Huwer}}, \bibinfo {author} {\bibfnamefont
  {J.}~\bibnamefont {Ghosh}}, \bibinfo {author} {\bibfnamefont
  {A.}~\bibnamefont {Haase}}, \bibinfo {author} {\bibfnamefont
  {M.}~\bibnamefont {Hennrich}}, \bibinfo {author} {\bibfnamefont
  {F.}~\bibnamefont {Dubin}}, \ and\ \bibinfo {author} {\bibfnamefont
  {J.}~\bibnamefont {Eschner}},\ }\href {http://dx.doi.org/10.1038/nphys1805}
  {\bibfield  {journal} {\bibinfo  {journal} {Nat Phys}\ }\textbf {\bibinfo
  {volume} {7}},\ \bibinfo {pages} {17} (\bibinfo {year} {2011})}\BibitemShut
  {NoStop}%
\bibitem [{\citenamefont {Maser}\ \emph {et~al.}(2016)\citenamefont {Maser},
  \citenamefont {Gmeiner}, \citenamefont {Utikal}, \citenamefont
  {G\"otzinger},\ and\ \citenamefont {Sandoghdar}}]{Maser2016}%
  \BibitemOpen
  \bibfield  {author} {\bibinfo {author} {\bibfnamefont {A.}~\bibnamefont
  {Maser}}, \bibinfo {author} {\bibfnamefont {B.}~\bibnamefont {Gmeiner}},
  \bibinfo {author} {\bibfnamefont {T.}~\bibnamefont {Utikal}}, \bibinfo
  {author} {\bibfnamefont {S.}~\bibnamefont {G\"otzinger}}, \ and\ \bibinfo
  {author} {\bibfnamefont {V.}~\bibnamefont {Sandoghdar}},\ }\href
  {http://dx.doi.org/10.1038/nphoton.2016.63} {\bibfield  {journal} {\bibinfo
  {journal} {Nat Photon}\ }\textbf {\bibinfo {volume} {10}},\ \bibinfo {pages}
  {450} (\bibinfo {year} {2016})}\BibitemShut {NoStop}%
\bibitem [{\citenamefont {Tey}\ \emph {et~al.}(2009)\citenamefont {Tey},
  \citenamefont {Maslennikov}, \citenamefont {Liew}, \citenamefont {Aljunid},
  \citenamefont {Huber}, \citenamefont {Chng}, \citenamefont {Chen},
  \citenamefont {Scarani},\ and\ \citenamefont {Kurtsiefer}}]{Tey2009}%
  \BibitemOpen
  \bibfield  {author} {\bibinfo {author} {\bibfnamefont {M.~K.}\ \bibnamefont
  {Tey}}, \bibinfo {author} {\bibfnamefont {G.}~\bibnamefont {Maslennikov}},
  \bibinfo {author} {\bibfnamefont {T.~C.~H.}\ \bibnamefont {Liew}}, \bibinfo
  {author} {\bibfnamefont {S.~A.}\ \bibnamefont {Aljunid}}, \bibinfo {author}
  {\bibfnamefont {F.}~\bibnamefont {Huber}}, \bibinfo {author} {\bibfnamefont
  {B.}~\bibnamefont {Chng}}, \bibinfo {author} {\bibfnamefont {Z.}~\bibnamefont
  {Chen}}, \bibinfo {author} {\bibfnamefont {V.}~\bibnamefont {Scarani}}, \
  and\ \bibinfo {author} {\bibfnamefont {C.}~\bibnamefont {Kurtsiefer}},\
  }\href {http://stacks.iop.org/1367-2630/11/i=4/a=043011} {\bibfield
  {journal} {\bibinfo  {journal} {New Journal of Physics}\ }\textbf {\bibinfo
  {volume} {11}},\ \bibinfo {pages} {043011} (\bibinfo {year}
  {2009})}\BibitemShut {NoStop}%
\bibitem [{\citenamefont {Hell}\ and\ \citenamefont
  {Stelzer}(1992)}]{Hell1992}%
  \BibitemOpen
  \bibfield  {author} {\bibinfo {author} {\bibfnamefont {S.}~\bibnamefont
  {Hell}}\ and\ \bibinfo {author} {\bibfnamefont {E.~H.~K.}\ \bibnamefont
  {Stelzer}},\ }\href {\doibase 10.1364/JOSAA.9.002159} {\bibfield  {journal}
  {\bibinfo  {journal} {J. Opt. Soc. Am. A}\ }\textbf {\bibinfo {volume} {9}},\
  \bibinfo {pages} {2159} (\bibinfo {year} {1992})}\BibitemShut {NoStop}%
\bibitem [{\citenamefont {Sondermann}\ \emph {et~al.}(2007)\citenamefont
  {Sondermann}, \citenamefont {Maiwald}, \citenamefont {Konermann},
  \citenamefont {Lindlein}, \citenamefont {Peschel},\ and\ \citenamefont
  {Leuchs}}]{Sondermann2007}%
  \BibitemOpen
  \bibfield  {author} {\bibinfo {author} {\bibfnamefont {M.}~\bibnamefont
  {Sondermann}}, \bibinfo {author} {\bibfnamefont {R.}~\bibnamefont {Maiwald}},
  \bibinfo {author} {\bibfnamefont {H.}~\bibnamefont {Konermann}}, \bibinfo
  {author} {\bibfnamefont {N.}~\bibnamefont {Lindlein}}, \bibinfo {author}
  {\bibfnamefont {U.}~\bibnamefont {Peschel}}, \ and\ \bibinfo {author}
  {\bibfnamefont {G.}~\bibnamefont {Leuchs}},\ }\href {\doibase
  10.1007/s00340-007-2859-4} {\bibfield  {journal} {\bibinfo  {journal}
  {Applied Physics B}\ }\textbf {\bibinfo {volume} {89}},\ \bibinfo {pages}
  {489} (\bibinfo {year} {2007})}\BibitemShut {NoStop}%
\bibitem [{\citenamefont {Leuchs}\ and\ \citenamefont
  {Sondermann}(2012)}]{Leuchs2012}%
  \BibitemOpen
  \bibfield  {author} {\bibinfo {author} {\bibfnamefont {G.}~\bibnamefont
  {Leuchs}}\ and\ \bibinfo {author} {\bibfnamefont {M.}~\bibnamefont
  {Sondermann}},\ }\href {http://www.ncbi.nlm.nih.gov/pmc/articles/PMC3627204/}
  {\bibfield  {journal} {\bibinfo  {journal} {Journal of Modern Optics}\
  }\textbf {\bibinfo {volume} {60}},\ \bibinfo {pages} {36} (\bibinfo {year}
  {2012})}\BibitemShut {NoStop}%
\bibitem [{\citenamefont {Abbe}(1873)}]{Abbe1873}%
  \BibitemOpen
  \bibfield  {author} {\bibinfo {author} {\bibfnamefont {E.}~\bibnamefont
  {Abbe}},\ }\href {\doibase 10.1007/BF02956173} {\bibfield  {journal}
  {\bibinfo  {journal} {Archiv f{\"u}r mikroskopische Anatomie}\ }\textbf
  {\bibinfo {volume} {9}},\ \bibinfo {pages} {413} (\bibinfo {year}
  {1873})}\BibitemShut {NoStop}%
\bibitem [{\citenamefont {Golla}\ \emph {et~al.}(2012)\citenamefont {Golla},
  \citenamefont {Chalopin}, \citenamefont {Bader}, \citenamefont {Harder},
  \citenamefont {Mantel}, \citenamefont {Maiwald}, \citenamefont {Lindlein},
  \citenamefont {Sondermann},\ and\ \citenamefont {Leuchs}}]{Golla2012}%
  \BibitemOpen
  \bibfield  {author} {\bibinfo {author} {\bibfnamefont {A.}~\bibnamefont
  {Golla}}, \bibinfo {author} {\bibfnamefont {B.}~\bibnamefont {Chalopin}},
  \bibinfo {author} {\bibfnamefont {M.}~\bibnamefont {Bader}}, \bibinfo
  {author} {\bibfnamefont {I.}~\bibnamefont {Harder}}, \bibinfo {author}
  {\bibfnamefont {K.}~\bibnamefont {Mantel}}, \bibinfo {author} {\bibfnamefont
  {R.}~\bibnamefont {Maiwald}}, \bibinfo {author} {\bibfnamefont
  {N.}~\bibnamefont {Lindlein}}, \bibinfo {author} {\bibfnamefont
  {M.}~\bibnamefont {Sondermann}}, \ and\ \bibinfo {author} {\bibfnamefont
  {G.}~\bibnamefont {Leuchs}},\ }\href {\doibase 10.1140/epjd/e2012-30293-y}
  {\bibfield  {journal} {\bibinfo  {journal} {The European Physical Journal D}\
  }\textbf {\bibinfo {volume} {66}},\ \bibinfo {pages} {190} (\bibinfo {year}
  {2012})}\BibitemShut {NoStop}%
\bibitem [{\citenamefont {Schlosser}\ \emph {et~al.}(2001)\citenamefont
  {Schlosser}, \citenamefont {Reymond}, \citenamefont {Protsenko},\ and\
  \citenamefont {Grangier}}]{Schlosser2001}%
  \BibitemOpen
  \bibfield  {author} {\bibinfo {author} {\bibfnamefont {N.}~\bibnamefont
  {Schlosser}}, \bibinfo {author} {\bibfnamefont {G.}~\bibnamefont {Reymond}},
  \bibinfo {author} {\bibfnamefont {I.}~\bibnamefont {Protsenko}}, \ and\
  \bibinfo {author} {\bibfnamefont {P.}~\bibnamefont {Grangier}},\ }\href
  {http://dx.doi.org/10.1038/35082512} {\bibfield  {journal} {\bibinfo
  {journal} {Nature}\ }\textbf {\bibinfo {volume} {411}},\ \bibinfo {pages}
  {1024} (\bibinfo {year} {2001})}\BibitemShut {NoStop}%
\bibitem [{\citenamefont {Chin}\ \emph {et~al.}(2017)\citenamefont {Chin},
  \citenamefont {Steiner},\ and\ \citenamefont {Kurtsiefer}}]{Chin2017}%
  \BibitemOpen
  \bibfield  {author} {\bibinfo {author} {\bibfnamefont {Y.-S.}\ \bibnamefont
  {Chin}}, \bibinfo {author} {\bibfnamefont {M.}~\bibnamefont {Steiner}}, \
  and\ \bibinfo {author} {\bibfnamefont {C.}~\bibnamefont {Kurtsiefer}},\
  }\href {\doibase 10.1103/PhysRevA.95.043809} {\bibfield  {journal} {\bibinfo
  {journal} {Phys. Rev. A}\ }\textbf {\bibinfo {volume} {95}},\ \bibinfo
  {pages} {043809} (\bibinfo {year} {2017})}\BibitemShut {NoStop}%
\bibitem [{\citenamefont {Slodi\ifmmode~\check{c}\else \v{c}\fi{}ka}\ \emph
  {et~al.}(2010)\citenamefont {Slodi\ifmmode~\check{c}\else \v{c}\fi{}ka},
  \citenamefont {H\'etet}, \citenamefont {Gerber}, \citenamefont {Hennrich},\
  and\ \citenamefont {Blatt}}]{Slodifmmodeheckclsecika2010}%
  \BibitemOpen
  \bibfield  {author} {\bibinfo {author} {\bibfnamefont {L.}~\bibnamefont
  {Slodi\ifmmode~\check{c}\else \v{c}\fi{}ka}}, \bibinfo {author}
  {\bibfnamefont {G.}~\bibnamefont {H\'etet}}, \bibinfo {author} {\bibfnamefont
  {S.}~\bibnamefont {Gerber}}, \bibinfo {author} {\bibfnamefont
  {M.}~\bibnamefont {Hennrich}}, \ and\ \bibinfo {author} {\bibfnamefont
  {R.}~\bibnamefont {Blatt}},\ }\href {\doibase 10.1103/PhysRevLett.105.153604}
  {\bibfield  {journal} {\bibinfo  {journal} {Phys. Rev. Lett.}\ }\textbf
  {\bibinfo {volume} {105}},\ \bibinfo {pages} {153604} (\bibinfo {year}
  {2010})}\BibitemShut {NoStop}%
\bibitem [{\citenamefont {Shen}\ and\ \citenamefont {Fan}(2007)}]{Shen2007}%
  \BibitemOpen
  \bibfield  {author} {\bibinfo {author} {\bibfnamefont {J.-T.}\ \bibnamefont
  {Shen}}\ and\ \bibinfo {author} {\bibfnamefont {S.}~\bibnamefont {Fan}},\
  }\href {\doibase 10.1103/PhysRevLett.98.153003} {\bibfield  {journal}
  {\bibinfo  {journal} {Phys. Rev. Lett.}\ }\textbf {\bibinfo {volume} {98}},\
  \bibinfo {pages} {153003} (\bibinfo {year} {2007})}\BibitemShut {NoStop}%
\bibitem [{\citenamefont {Birnbaum}\ \emph {et~al.}(2005)\citenamefont
  {Birnbaum}, \citenamefont {Boca}, \citenamefont {Miller}, \citenamefont
  {Boozer}, \citenamefont {Northup},\ and\ \citenamefont
  {Kimble}}]{Birnbaum2005}%
  \BibitemOpen
  \bibfield  {author} {\bibinfo {author} {\bibfnamefont {K.~M.}\ \bibnamefont
  {Birnbaum}}, \bibinfo {author} {\bibfnamefont {A.}~\bibnamefont {Boca}},
  \bibinfo {author} {\bibfnamefont {R.}~\bibnamefont {Miller}}, \bibinfo
  {author} {\bibfnamefont {A.~D.}\ \bibnamefont {Boozer}}, \bibinfo {author}
  {\bibfnamefont {T.~E.}\ \bibnamefont {Northup}}, \ and\ \bibinfo {author}
  {\bibfnamefont {H.~J.}\ \bibnamefont {Kimble}},\ }\href
  {http://dx.doi.org/10.1038/nature03804} {\bibfield  {journal} {\bibinfo
  {journal} {Nature}\ }\textbf {\bibinfo {volume} {436}},\ \bibinfo {pages}
  {87} (\bibinfo {year} {2005})}\BibitemShut {NoStop}%
\bibitem [{\citenamefont {Dayan}\ \emph {et~al.}(2008)\citenamefont {Dayan},
  \citenamefont {Parkins}, \citenamefont {Aoki}, \citenamefont {Ostby},
  \citenamefont {Vahala},\ and\ \citenamefont {Kimble}}]{Dayan2008}%
  \BibitemOpen
  \bibfield  {author} {\bibinfo {author} {\bibfnamefont {B.}~\bibnamefont
  {Dayan}}, \bibinfo {author} {\bibfnamefont {A.~S.}\ \bibnamefont {Parkins}},
  \bibinfo {author} {\bibfnamefont {T.}~\bibnamefont {Aoki}}, \bibinfo {author}
  {\bibfnamefont {E.~P.}\ \bibnamefont {Ostby}}, \bibinfo {author}
  {\bibfnamefont {K.~J.}\ \bibnamefont {Vahala}}, \ and\ \bibinfo {author}
  {\bibfnamefont {H.~J.}\ \bibnamefont {Kimble}},\ }\href {\doibase
  10.1126/science.1152261} {\bibfield  {journal} {\bibinfo  {journal}
  {Science}\ }\textbf {\bibinfo {volume} {319}},\ \bibinfo {pages} {1062}
  (\bibinfo {year} {2008})}\BibitemShut {NoStop}%
\bibitem [{\citenamefont {Reinhard}\ \emph {et~al.}(2012)\citenamefont
  {Reinhard}, \citenamefont {Volz}, \citenamefont {Winger}, \citenamefont
  {Badolato}, \citenamefont {Hennessy}, \citenamefont {Hu},\ and\ \citenamefont
  {Imamo{\u{g}}lu}}]{Reinhard2012}%
  \BibitemOpen
  \bibfield  {author} {\bibinfo {author} {\bibfnamefont {A.}~\bibnamefont
  {Reinhard}}, \bibinfo {author} {\bibfnamefont {T.}~\bibnamefont {Volz}},
  \bibinfo {author} {\bibfnamefont {M.}~\bibnamefont {Winger}}, \bibinfo
  {author} {\bibfnamefont {A.}~\bibnamefont {Badolato}}, \bibinfo {author}
  {\bibfnamefont {K.~J.}\ \bibnamefont {Hennessy}}, \bibinfo {author}
  {\bibfnamefont {E.~L.}\ \bibnamefont {Hu}}, \ and\ \bibinfo {author}
  {\bibfnamefont {A.}~\bibnamefont {Imamo{\u{g}}lu}},\ }\href
  {http://dx.doi.org/10.1038/nphoton.2011.321} {\bibfield  {journal} {\bibinfo
  {journal} {Nat Photon}\ }\textbf {\bibinfo {volume} {6}},\ \bibinfo {pages}
  {93} (\bibinfo {year} {2012})}\BibitemShut {NoStop}%
\bibitem [{\citenamefont {Hoi}\ \emph {et~al.}(2012)\citenamefont {Hoi},
  \citenamefont {Palomaki}, \citenamefont {Lindkvist}, \citenamefont
  {Johansson}, \citenamefont {Delsing},\ and\ \citenamefont
  {Wilson}}]{Hoi2012}%
  \BibitemOpen
  \bibfield  {author} {\bibinfo {author} {\bibfnamefont {I.-C.}\ \bibnamefont
  {Hoi}}, \bibinfo {author} {\bibfnamefont {T.}~\bibnamefont {Palomaki}},
  \bibinfo {author} {\bibfnamefont {J.}~\bibnamefont {Lindkvist}}, \bibinfo
  {author} {\bibfnamefont {G.}~\bibnamefont {Johansson}}, \bibinfo {author}
  {\bibfnamefont {P.}~\bibnamefont {Delsing}}, \ and\ \bibinfo {author}
  {\bibfnamefont {C.~M.}\ \bibnamefont {Wilson}},\ }\href {\doibase
  10.1103/PhysRevLett.108.263601} {\bibfield  {journal} {\bibinfo  {journal}
  {Phys. Rev. Lett.}\ }\textbf {\bibinfo {volume} {108}},\ \bibinfo {pages}
  {263601} (\bibinfo {year} {2012})}\BibitemShut {NoStop}%
\bibitem [{\citenamefont {Peyronel}\ \emph {et~al.}(2012)\citenamefont
  {Peyronel}, \citenamefont {Firstenberg}, \citenamefont {Liang}, \citenamefont
  {Hofferberth}, \citenamefont {Gorshkov}, \citenamefont {Pohl}, \citenamefont
  {Lukin},\ and\ \citenamefont {Vuletic}}]{Peyronel2012}%
  \BibitemOpen
  \bibfield  {author} {\bibinfo {author} {\bibfnamefont {T.}~\bibnamefont
  {Peyronel}}, \bibinfo {author} {\bibfnamefont {O.}~\bibnamefont
  {Firstenberg}}, \bibinfo {author} {\bibfnamefont {Q.-Y.}\ \bibnamefont
  {Liang}}, \bibinfo {author} {\bibfnamefont {S.}~\bibnamefont {Hofferberth}},
  \bibinfo {author} {\bibfnamefont {A.~V.}\ \bibnamefont {Gorshkov}}, \bibinfo
  {author} {\bibfnamefont {T.}~\bibnamefont {Pohl}}, \bibinfo {author}
  {\bibfnamefont {M.~D.}\ \bibnamefont {Lukin}}, \ and\ \bibinfo {author}
  {\bibfnamefont {V.}~\bibnamefont {Vuletic}},\ }\href
  {http://dx.doi.org/10.1038/nature11361} {\bibfield  {journal} {\bibinfo
  {journal} {Nature}\ }\textbf {\bibinfo {volume} {488}},\ \bibinfo {pages}
  {57} (\bibinfo {year} {2012})}\BibitemShut {NoStop}%
\bibitem [{\citenamefont {Chang}\ \emph {et~al.}(2007)\citenamefont {Chang},
  \citenamefont {Sorensen}, \citenamefont {Demler},\ and\ \citenamefont
  {Lukin}}]{Chang2007}%
  \BibitemOpen
  \bibfield  {author} {\bibinfo {author} {\bibfnamefont {D.~E.}\ \bibnamefont
  {Chang}}, \bibinfo {author} {\bibfnamefont {A.~S.}\ \bibnamefont {Sorensen}},
  \bibinfo {author} {\bibfnamefont {E.~A.}\ \bibnamefont {Demler}}, \ and\
  \bibinfo {author} {\bibfnamefont {M.~D.}\ \bibnamefont {Lukin}},\ }\href
  {http://dx.doi.org/10.1038/nphys708} {\bibfield  {journal} {\bibinfo
  {journal} {Nat Phys}\ }\textbf {\bibinfo {volume} {3}},\ \bibinfo {pages}
  {807} (\bibinfo {year} {2007})}\BibitemShut {NoStop}%
\bibitem [{\citenamefont {Zheng}\ \emph {et~al.}(2010)\citenamefont {Zheng},
  \citenamefont {Gauthier},\ and\ \citenamefont {Baranger}}]{Zheng2010}%
  \BibitemOpen
  \bibfield  {author} {\bibinfo {author} {\bibfnamefont {H.}~\bibnamefont
  {Zheng}}, \bibinfo {author} {\bibfnamefont {D.~J.}\ \bibnamefont {Gauthier}},
  \ and\ \bibinfo {author} {\bibfnamefont {H.~U.}\ \bibnamefont {Baranger}},\
  }\href {\doibase 10.1103/PhysRevA.82.063816} {\bibfield  {journal} {\bibinfo
  {journal} {Phys. Rev. A}\ }\textbf {\bibinfo {volume} {82}},\ \bibinfo
  {pages} {063816} (\bibinfo {year} {2010})}\BibitemShut {NoStop}%
\bibitem [{\citenamefont {Gomer}\ \emph {et~al.}(1998)\citenamefont {Gomer},
  \citenamefont {Ueberholz}, \citenamefont {Knappe}, \citenamefont {Strauch},
  \citenamefont {Frese},\ and\ \citenamefont {Meschede}}]{Gomer1998}%
  \BibitemOpen
  \bibfield  {author} {\bibinfo {author} {\bibfnamefont {V.}~\bibnamefont
  {Gomer}}, \bibinfo {author} {\bibfnamefont {B.}~\bibnamefont {Ueberholz}},
  \bibinfo {author} {\bibfnamefont {S.}~\bibnamefont {Knappe}}, \bibinfo
  {author} {\bibfnamefont {F.}~\bibnamefont {Strauch}}, \bibinfo {author}
  {\bibfnamefont {D.}~\bibnamefont {Frese}}, \ and\ \bibinfo {author}
  {\bibfnamefont {D.}~\bibnamefont {Meschede}},\ }\href {\doibase
  10.1007/s003400050567} {\bibfield  {journal} {\bibinfo  {journal} {Applied
  Physics B}\ }\textbf {\bibinfo {volume} {67}},\ \bibinfo {pages} {689}
  (\bibinfo {year} {1998})}\BibitemShut {NoStop}%
\bibitem [{\citenamefont {Weber}\ \emph {et~al.}(2006)\citenamefont {Weber},
  \citenamefont {Volz}, \citenamefont {Saucke}, \citenamefont {Kurtsiefer},\
  and\ \citenamefont {Weinfurter}}]{Weber2006}%
  \BibitemOpen
  \bibfield  {author} {\bibinfo {author} {\bibfnamefont {M.}~\bibnamefont
  {Weber}}, \bibinfo {author} {\bibfnamefont {J.}~\bibnamefont {Volz}},
  \bibinfo {author} {\bibfnamefont {K.}~\bibnamefont {Saucke}}, \bibinfo
  {author} {\bibfnamefont {C.}~\bibnamefont {Kurtsiefer}}, \ and\ \bibinfo
  {author} {\bibfnamefont {H.}~\bibnamefont {Weinfurter}},\ }\href {\doibase
  10.1103/PhysRevA.73.043406} {\bibfield  {journal} {\bibinfo  {journal} {Phys.
  Rev. A}\ }\textbf {\bibinfo {volume} {73}},\ \bibinfo {pages} {043406}
  (\bibinfo {year} {2006})}\BibitemShut {NoStop}%
\bibitem [{\citenamefont {Chang}\ \emph {et~al.}(2008)\citenamefont {Chang},
  \citenamefont {Gritsev}, \citenamefont {Morigi}, \citenamefont {Vuletic},
  \citenamefont {Lukin},\ and\ \citenamefont {Demler}}]{Chang2008}%
  \BibitemOpen
  \bibfield  {author} {\bibinfo {author} {\bibfnamefont {D.~E.}\ \bibnamefont
  {Chang}}, \bibinfo {author} {\bibfnamefont {V.}~\bibnamefont {Gritsev}},
  \bibinfo {author} {\bibfnamefont {G.}~\bibnamefont {Morigi}}, \bibinfo
  {author} {\bibfnamefont {V.}~\bibnamefont {Vuletic}}, \bibinfo {author}
  {\bibfnamefont {M.~D.}\ \bibnamefont {Lukin}}, \ and\ \bibinfo {author}
  {\bibfnamefont {E.~A.}\ \bibnamefont {Demler}},\ }\href
  {http://dx.doi.org/10.1038/nphys1074} {\bibfield  {journal} {\bibinfo
  {journal} {Nat Phys}\ }\textbf {\bibinfo {volume} {4}},\ \bibinfo {pages}
  {884} (\bibinfo {year} {2008})}\BibitemShut {NoStop}%
\bibitem [{\citenamefont {Chang}\ \emph {et~al.}(2014)\citenamefont {Chang},
  \citenamefont {Vuletic},\ and\ \citenamefont {Lukin}}]{Chang2014}%
  \BibitemOpen
  \bibfield  {author} {\bibinfo {author} {\bibfnamefont {D.~E.}\ \bibnamefont
  {Chang}}, \bibinfo {author} {\bibfnamefont {V.}~\bibnamefont {Vuletic}}, \
  and\ \bibinfo {author} {\bibfnamefont {M.~D.}\ \bibnamefont {Lukin}},\ }\href
  {http://dx.doi.org/10.1038/nphoton.2014.192} {\bibfield  {journal} {\bibinfo
  {journal} {Nat Photon}\ }\textbf {\bibinfo {volume} {8}},\ \bibinfo {pages}
  {685} (\bibinfo {year} {2014})}\BibitemShut {NoStop}%
\bibitem [{\citenamefont {Reiserer}\ and\ \citenamefont
  {Rempe}(2015)}]{Reiserer2015}%
  \BibitemOpen
  \bibfield  {author} {\bibinfo {author} {\bibfnamefont {A.}~\bibnamefont
  {Reiserer}}\ and\ \bibinfo {author} {\bibfnamefont {G.}~\bibnamefont
  {Rempe}},\ }\href {\doibase 10.1103/RevModPhys.87.1379} {\bibfield  {journal}
  {\bibinfo  {journal} {Rev. Mod. Phys.}\ }\textbf {\bibinfo {volume} {87}},\
  \bibinfo {pages} {1379} (\bibinfo {year} {2015})}\BibitemShut {NoStop}%
\bibitem [{\citenamefont {Pritchard}\ \emph {et~al.}(2010)\citenamefont
  {Pritchard}, \citenamefont {Maxwell}, \citenamefont {Gauguet}, \citenamefont
  {Weatherill}, \citenamefont {Jones},\ and\ \citenamefont
  {Adams}}]{Pritchard2010}%
  \BibitemOpen
  \bibfield  {author} {\bibinfo {author} {\bibfnamefont {J.~D.}\ \bibnamefont
  {Pritchard}}, \bibinfo {author} {\bibfnamefont {D.}~\bibnamefont {Maxwell}},
  \bibinfo {author} {\bibfnamefont {A.}~\bibnamefont {Gauguet}}, \bibinfo
  {author} {\bibfnamefont {K.~J.}\ \bibnamefont {Weatherill}}, \bibinfo
  {author} {\bibfnamefont {M.~P.~A.}\ \bibnamefont {Jones}}, \ and\ \bibinfo
  {author} {\bibfnamefont {C.~S.}\ \bibnamefont {Adams}},\ }\href {\doibase
  10.1103/PhysRevLett.105.193603} {\bibfield  {journal} {\bibinfo  {journal}
  {Phys. Rev. Lett.}\ }\textbf {\bibinfo {volume} {105}},\ \bibinfo {pages}
  {193603} (\bibinfo {year} {2010})}\BibitemShut {NoStop}%
\bibitem [{\citenamefont {Firstenberg}\ \emph {et~al.}(2016)\citenamefont
  {Firstenberg}, \citenamefont {Adams},\ and\ \citenamefont
  {Hofferberth}}]{Firstenberg2016}%
  \BibitemOpen
  \bibfield  {author} {\bibinfo {author} {\bibfnamefont {O.}~\bibnamefont
  {Firstenberg}}, \bibinfo {author} {\bibfnamefont {C.~S.}\ \bibnamefont
  {Adams}}, \ and\ \bibinfo {author} {\bibfnamefont {S.}~\bibnamefont
  {Hofferberth}},\ }\href {http://stacks.iop.org/0953-4075/49/i=15/a=152003}
  {\bibfield  {journal} {\bibinfo  {journal} {Journal of Physics B: Atomic,
  Molecular and Optical Physics}\ }\textbf {\bibinfo {volume} {49}},\ \bibinfo
  {pages} {152003} (\bibinfo {year} {2016})}\BibitemShut {NoStop}%
\end{thebibliography}
%merlin.mbs apsrev4-1.bst 2010-07-25 4.21a (PWD, AO, DPC) hacked
%Control: key (0)
%Control: author (72) initials jnrlst
%Control: editor formatted (1) identically to author
%Control: production of article title (-1) disabled
%Control: page (0) single
%Control: year (1) truncated
%Control: production of eprint (0) enabled
%

\end{document}